\documentclass[aps,prd,twocolumn,superscriptaddress,nofootinbib,
    colorlinks=true, linkcolor=magenta, citecolor=magenta, urlcolor=magenta]{revtex4-2}
\usepackage{epsfig}
\usepackage{amsmath}
\usepackage{amsfonts}
\usepackage{float}
\usepackage{amssymb}
\usepackage{slashed}
\usepackage[dvipsnames]{xcolor}
\usepackage{pbox}
\usepackage{subfigure}
\usepackage{tabularx}
\usepackage{titlesec}
\usepackage{scrextend}
\usepackage{enumitem}
\usepackage[normalem]{ulem}
\usepackage{orcidlink}
\usepackage{natbib}
\usepackage{xcolor}
\usepackage{braket}

\graphicspath{{Figures/}}
\def\bal#1\eal{\begin{align}#1\end{align}}
\def\bea#1\eea{\begin{eqnarray}#1\end{eqnarray}}
\def\beq#1\eeq{\begin{equation}#1\end{equation}}

\begin{document}
\hypersetup{colorlinks=true, citecolor=magenta, linkcolor=magenta, urlcolor=magenta}

\title{Axion–Photon Conversion in Variable-Domain Magnetic Fields and the Highest-Energy Photons from GRB 221009A}

\author{Rick Sundar Garai\orcidlink{0009-0008-9357-2257}}
 \email{ricksundargarai@gmail.com}
 \affiliation{School of Physical Sciences, Indian Association for the Cultivation of Science, 2A \& 2B Raja S.C. Mullick Road, Jadavpur, Kolkata 700032, India}

\author{Tanmoy Kumar\orcidlink{0000-0001-9775-6645}}
 \email{kumartanmoy1998@gmail.com}
 \affiliation{Department of Physics, Indian Institute of Technology, Kanpur 208016, India}

\author{Sourov Roy\orcidlink{0000-0002-1015-3241}}
 \email{tpsr@iacs.res.in}
 \affiliation{School of Physical Sciences, Indian Association for the Cultivation of Science, 2A \& 2B Raja S.C. Mullick Road, Jadavpur, Kolkata 700032, India}

\begin{abstract}

We investigate photon--ALP conversion as a simultaneous explanation of the ultra-high-energy photons observed from GRB~221009A by LHAASO and of the $\sim300$~TeV photon reported by Carpet-3. We model the magnetic field of the host galaxy as a set of domains with randomly oriented field directions and variable domain lengths, and propagate the photon-ALP system through the host galaxy, the intergalactic medium and the Milky Way. Fitting the combined LHAASO and Carpet-3 spectra, we find that a broad distribution of domain lengths raises the photon survival probability at the highest energies and accommodates the Carpet-3 point with photon--ALP conversion alone. The best fit, $m_a=3.85\times10^{-7}$~eV and $g_{a\gamma}=3.07\times10^{-11}\,\mathrm{GeV}^{-1}$, lies below the CAST bound, and parts of the $1\sigma$ and $2\sigma$ regions are also compatible with it. 

\end{abstract}

\maketitle
\section{Introduction}

Axions and axion-like particles (ALPs) are among the most well-motivated extensions of the Standard Model, arising generically as pseudo-Nambu-Goldstone bosons (pNGBs) of a spontaneously broken global chiral $U(1)$ symmetry~\cite{Marsh:2015xka,Irastorza:2018dyq,Choi:2020rgn}. While the QCD axion initially arose as a by-product of the Peccei-Quinn solution to the strong CP problem~\cite{Peccei:1977hh,Peccei:1977ur,Weinberg:1977ma,Wilczek:1977pj,Peccei:2006as}, a much broader class of ALPs emerge naturally in string-theoretic constructions and other ultraviolet completions~\cite{Arvanitaki:2009fg,Svrcek:2006yi}, where they need not carry any relation to the QCD scale. Consequently, ALPs\footnote{In this work we use the terms axions and ALPs interchangeably to refer to generic axion-like particles.} span an enormous range of viable masses, from $\sim 10^{-20}$ eV to several GeV, and couple only extremely weakly to Standard Model fields, thereby also making them good candidates for dark matter~\cite{Preskill:1982cy,Abbott:1982af,Dine:1982ah,Khlopov:1999tm}. 
A generic and phenomenologically central feature of these particles is their coupling to two photons, parametrized by the coupling constant $g_{a\gamma}$~\cite{Kim:2008hd,DiLuzio:2020wdo}. This coupling gives rise to two phenomenologically important processes: (a) in the presence of an external magnetic field, the photon can convert into an axion and vice versa, leading to axion--photon oscillation~\cite{Sikivie:1983ip,Raffelt:1987im}, and (b) an axion with sufficient mass can decay into two photons~\cite{DiLuzio:2020wdo,Cadamuro:2011fd,Caputo:2024oqc}. These two processes serve as the basis of the majority of the astrophysical searches for axions and ALPs.

Ultra-high-energy (UHE) photons with energies in the range of a few GeV to several TeV are produced in some of the most extreme astrophysical environments in the universe, such as active galactic nuclei (AGN), gamma-ray bursts (GRBs), and quasars~\cite{Madejski:2016abc,Padovani:2017zpf,Blandford:2018iot}. In these environments, which typically host strong magnetic fields~\cite{Pudritz:2012ssr,Kumar:2014upa}, charged particles are accelerated to ultra-relativistic energies and subsequently interact with the surrounding medium and radiation fields, producing UHE photons via processes such as synchrotron emission, inverse Compton scattering, and hadronic interactions~\cite{Meszaros:2006rc,Boettcher:2013wxa}.
UHE photons having energies above $1$ TeV, while traveling through the intergalactic medium, interact with the extragalactic background light (EBL) and get heavily attenuated. The dominant process for absorption of a photon of energy few TeV is electron-positron pair production $\gamma\gamma\rightarrow e^+e^-$~\cite{Nikishov:1962rmq, 1967PhRv..155.1404G}.
Hence the flux of UHE photons from distant sources is expected to be negligible because of this attenuation.

On 9 October 2022, an extremely powerful gamma-ray burst, GRB~221009A, located at RA $=288.26^\circ$, Dec $=+19.77^\circ$~\cite{Dichiara:2022abc,Burns:2023oxn} with redshift $z=0.151$~\cite{Malesani:2023aom}, was detected by several observatories. The Large High Altitude Air Shower Observatory (LHAASO) registered more than $5{,}000$ very-high-energy photons above $0.2$~TeV, with individual events extending up to $18$~TeV~\cite{LHAASO:2023tbl,LHAASA:2023pay}. Subsequently, a $251$~TeV photon-like air-shower event in spatial and temporal coincidence with the burst was reported by the Carpet-2 array~\cite{Carpet3:2022uhy}, later revised to $300$~TeV following reanalysis with the upgraded Carpet-3 detector~\cite{Carpet-3Group:2025fcs}, making it the highest-energy photon-like event ever associated with a GRB. Given the distance of the source from Earth, such a UHE photon should have been attenuated before reaching the Earth. Hence the observation of this UHE photon is a mystery.

UHE photons from extragalactic sources can undergo photon--ALP mixing in the presence of the magnetic field near the source galaxy, converting a fraction of the flux into a photon--ALP beam~\cite{Fairbairn:2009zi,Galanti:2022ijh}. While propagating through the intergalactic medium, the photon component continues to suffer EBL attenuation as described above, whereas the axion component, being weakly interacting with the background radiation fields, propagates unimpeded. The surviving photons and the unattenuated axion flux thus reach our Milky Way together, where passage through the Galactic magnetic field allows a fraction of the axions to reconvert into photons. This photon$\to$ALP$\to$photon conversion can therefore replenish the UHE photon flux at Earth beyond what EBL attenuation alone would permit, offering a possible resolution to the puzzle described above.

There have been attempts to explain these anomalous observations through photon--ALP conversion. Refs.~\cite{Qin:2025olt,Ofengeim:2025jsw, Satunin:2025hbk, Finke:2022swf, Galanti:2022chk, PhysRevD.108.023002} argue that a pure photon--ALP oscillation scenario, assuming a fixed domain length for the source-galaxy magnetic field, cannot simultaneously account for the $\sim300$~TeV Carpet-3 event together with the LHAASO spectrum extending up to $\sim18$~TeV, and consequently invoke Lorentz invariance violation (LIV) in conjunction with photon--ALP mixing to accommodate the $300$~TeV photon. In this work however, we show that if one incorporates a variable magnetic field domain length in the source-galaxy, a purely ALP-based scenario, without invoking LIV, can accommodate both the LHAASO and Carpet-3 measurements {while still remaining outside the CAST excluded region}.

The rest of this paper is organized as follows. In Sec.~\ref{sec:formalism}, we review photon--ALP oscillations in uniform and fixed-length multi-domain magnetic fields, and extend the formalism to domains of variable length. The conversion probability then follows from a Gaussian average of the fixed-domain result over the domain length, and the standard deviation of this distribution enters as an additional term that raises the conversion probability. In Sec.~\ref{sec:grb}, we summarize the GRB~221009A observations and the standard EBL attenuation of the photon flux. In Sec.~\ref{sec:photon_prop}, we include photon--ALP conversion in the source, the intergalactic medium and the Milky Way, and describe the fit to the LHAASO and Carpet-3 spectra. In Sec.~\ref{sec:results}, we present the best-fit $m_a$ and $g_{a\gamma}$ for fixed and variable domain lengths, compare them with the CAST bound, and conclude.

\section{Axion--Photon Oscillation}
\label{sec:formalism}

The Lagrangian of the axion field relevant for our discussion is given by~\cite{Raffelt:1987im,DiLuzio:2020wdo}
\begin{equation}\label{Lagrangian}
  \mathcal{L}=-\frac{1}{4}F_{\mu\nu}F^{\mu\nu}+\frac{1}{2}\left(\partial_\mu a \partial^\mu a - m_a^2a^2\right)+g_{a\gamma}F_{\mu\nu}\tilde F^{\mu\nu}a\,,
\end{equation}
where $a$ is the axion field, $m_a$ its mass, $F_{\mu\nu}$ the electromagnetic field strength tensor, and $\tilde F^{\mu\nu} = \frac{1}{2}\epsilon^{\mu\nu\rho\sigma}F_{\rho\sigma}$ its dual.

Starting from the Lagrangian in Eq.~\eqref{Lagrangian}, we can derive the equations of motion for the axion and the photon fields as
\begin{equation}
(\Box - m_a^2)a=\dfrac{1}{4} g_{a\gamma}F_{\mu\nu}\tilde{F}^{\mu\nu}=-g_{a\gamma} \bf{E} \cdot \bf{B},
\end{equation}
and
\begin{equation}
\partial_\mu F^{\mu\nu}=-g_{a\gamma}\tilde{F}^{\mu\nu}\partial_\mu a.
\end{equation}

Working in the radiation gauge and following Ref.~\cite{Raffelt:1987im} we obtain the Schr\"odinger-like linearized equation of motion as
\begin{equation}
i\frac{d}{dz}\Psi
=
\mathcal{M}
\Psi
\label{eq:linearized_EOM}
\end{equation}
where we have defined

\begin{equation}
    \Psi=
\begin{pmatrix}
A_{\perp}(z)\\
A_{\parallel}(z)\\
a(z)
\end{pmatrix}
e^{-i\omega z}
\end{equation}
and
\begin{equation}
    \mathcal{M} =
\begin{pmatrix}
0 & 0 & 0
\\[10pt]
0 & 0 & -\dfrac{1}{2}g_{a\gamma}B_T
\\[10pt]
0 &-\dfrac{1}{2}g_{a\gamma}B_T & -\dfrac{m_a^2}{2 E_a}
\end{pmatrix}
\label{eq:vacuum_mixing}
\end{equation}
is the axion--photon mixing matrix in vacuum, where for simplicity we have assumed that the external magnetic field is transverse to the direction of propagation with $B_T$ being its magnitude and $E_a$ is the axion energy which is equal to the photon energy $E_\gamma$. The presence of the off-diagonal terms in the mixing matrix results in the axion and photon modes getting mixed as they propagate through a magnetized medium resulting in the conversion of axions into photons and vice-versa.

In realistic astrophysical environments, photon propagation is modified by the presence of background magnetic fields and plasma. Consequently, the axion--photon mixing matrix is of the form

\begin{equation}
    \mathcal{M} =
\begin{pmatrix}
\Delta_\perp & 0 & 0
\\[10pt]
0 & \Delta_\parallel & \Delta_{a \gamma}
\\[10pt]
0 &\Delta_{a \gamma} & \Delta_a
\end{pmatrix}
\label{eq:plasma_mixing}
\end{equation}

with $\Delta_{a \gamma} = -{1}/{2}g_{a\gamma}B_T$ and 
$\Delta_a = -{m_a^2}/{2 E_a}$.

The diagonal entries $\Delta_\parallel$ and $\Delta_\perp$ receive contributions from vacuum polarization and the plasma~\cite{Carenza:2021alz}
\begin{equation}
\Delta_* = \Delta_{\rm QED} + \Delta_{\rm plasma},
\qquad (*=\parallel,\perp).
\end{equation}
The plasma contribution is
\begin{equation}
\Delta_{\rm plasma} = -\frac{\omega_p^2}{2 E_a},
\end{equation}
where
\begin{equation}
\omega_p^2 = e^2\frac{n_e}{m_e},
\end{equation}
is the plasma frequency and $n_e$ is the ambient electron number density.

For the astrophysical systems considered in this work, the QED correction is much smaller than the plasma contribution and is therefore neglected.

With the magnetic field and electron density profiles specified, the mixing matrix in Eq.~\eqref{eq:plasma_mixing} can be constructed explicitly, and Eq.~\eqref{eq:linearized_EOM} solved to obtain the axion-to-photon conversion probability.

\subsection{Single Domain: Constant Magnitude and Direction}
\label{subsec:single-domain}

We first recall the standard single-domain result~\cite{Raffelt:1987im}, in which the magnetic field has constant magnitude $B_T$ and fixed direction over the full propagation path, so that the mixing term $\Delta_{a\gamma}$ is independent of $z$. In this simplest case, the axion mixes only with the parallel photon polarization $A_{\parallel}$, and the mixing matrix reduces to the $2\times2$ system
\begin{equation}\label{11}
i\frac{d}{dz}
\begin{pmatrix}
A_{\parallel}(z)\\
a(z)
\end{pmatrix}
=
\begin{pmatrix}
\Delta_{\parallel} & \Delta_{a\gamma}\\
\Delta_{a\gamma} & \Delta_{a}
\end{pmatrix}
\begin{pmatrix}
A_{\parallel}(z)\\
a(z)
\end{pmatrix}.
\end{equation}
Diagonalizing the mixing matrix in Eq.~\eqref{11} with mixing angle $\theta$, where
\begin{equation}
\tan2\theta = \frac{2\Delta_{a\gamma}}{\Delta_a-\Delta_{\parallel}},
\end{equation}
one obtains the photon-to-axion conversion probability for $a(0)=0$, $A_{\parallel}(0)=1$,
\begin{equation}\label{17}
P_0(\gamma\rightarrow a)
=
(\sin2\theta)^2
\sin^2\left(\frac{\Delta_{\rm osc}}{2}z\right),
\end{equation}
where the oscillation wavenumber is
\begin{equation}\label{oscillation_length}
\Delta_{\rm osc}
\equiv
\sqrt{(\Delta_a-\Delta_{\parallel})^2+(2\Delta_{a\gamma})^2}.
\end{equation}
Using $\sin2\theta = 2\Delta_{a\gamma}/\Delta_{\rm osc}$, Eq.~\eqref{17} can equivalently be written as
\begin{equation}\label{oscillation_eqn_single_domain}
P_0(\gamma\rightarrow a)
=
(\Delta_{a\gamma} z)^2
\frac{\sin^2\left(\dfrac{\Delta_{\rm osc}}{2}z\right)}
{\left(\dfrac{\Delta_{\rm osc}}{2}z\right)^2}.
\end{equation}
Conversion is most efficient in the strong-mixing regime $\Delta_{a\gamma}\gg|\Delta_a-\Delta_{\parallel}|$, where $\theta\sim\pi/4$ and $\Delta_{{\rm osc}}\sim2\Delta_{a\gamma}$, giving the probability
\begin{equation}\label{singledomain}
P_{0}(\gamma\rightarrow a)\sim\Delta_{a\gamma}^2 z^2.
\end{equation}

\subsection{Multi-Domain: Fixed Domain Length}
\label{subsec:multi-domain-random-direction}
Astrophysical magnetic fields are rarely uniform over the full path traversed by a photon or axion. In many environments of interest such as galaxy clusters, the intracluster and intergalactic medium, stellar and neutron star magnetospheres, the field is better described as coherent only over finite length scales, known as domains or cells, within which the magnitude and direction of the magnetic field can be treated as approximately constant. Beyond a single domain, the field is expected to change, either in direction, in magnitude, or both, reflecting the turbulent or structured nature of the underlying plasma. This domain-like structure has a direct impact on axion--photon conversion: since the mixing angle and oscillation length both depend on the local field configuration, a photon or axion traversing multiple domains experiences a series of distinct mixing conditions rather than a single coherent oscillation. 

We recall the simple multi-domain treatment~\cite{Grossman:2002by,Csaki:2001yk} in which photons traverse a series of $N$ domains of equal length $z_{\rm dom}$, in each of which the magnetic field is constant and homogeneous. The transverse component of the field is assumed to have the same magnitude in every domain, but a random orientation from one domain to the next, with domains separated by intervals over which conversion is negligible.

Consider an initial state
\begin{equation}
|\Psi(0)\rangle = c_1(0)|A_{\perp}\rangle + c_2(0)|A_{\parallel}\rangle + c_a(0)|a\rangle,
\end{equation}
where $|A_{\perp}\rangle$ and $|A_{\parallel}\rangle$ denote the photon polarization states perpendicular and parallel to the magnetic field in the first domain, so that the initial photon and axion fluxes are $I_{\gamma}^0\sim|c_1(0)|^2+|c_2(0)|^2$ and $I_a^0\sim|c_a(0)|^2$. In the $n$-th domain the field is rotated by an angle $\theta_n$ relative to the first domain, and the local polarization basis is related to the fixed basis by
\begin{equation}
|A_{\parallel}^{\,n}\rangle = c_n|A_{\perp}\rangle + s_n|A_{\parallel}\rangle,
\qquad
|A_{\perp}^{\,n}\rangle = -s_n|A_{\perp}\rangle + c_n|A_{\parallel}\rangle,
\end{equation}
with $c_n\equiv\cos\theta_n$, $s_n\equiv\sin\theta_n$.

Within each domain, the axion mixes only with the locally parallel polarization according to the mixing matrix of Eq.~\eqref{11}, with transition amplitudes $t_{\gamma\rightarrow\gamma}$ and $t_{\gamma\rightarrow a}$ that are identical in every domain, since the field magnitude is fixed. The single-domain conversion probability is therefore $P_0=|t_{\gamma\rightarrow a}|^2$, given by Eq.~\eqref{17}, with photon survival probability $1-P_0$. Propagating the fluxes domain-by-domain and averaging over the random orientations $\theta_n$ -- so that $c_n^2,s_n^2\to1/2$ and $c_n$, $s_n$, $c_ns_n\to0$, gives the recursion relation
\begin{equation}\label{matrix_multiply}
\begin{pmatrix}
I_{\gamma}^{n+1}\\
I_a^{n+1}
\end{pmatrix}
=
\begin{pmatrix}
1-\dfrac{1}{2}P_0 & P_0\\[6pt]
\dfrac{1}{2}P_0 & 1-P_0
\end{pmatrix}
\begin{pmatrix}
I_{\gamma}^{n}\\
I_a^{n}
\end{pmatrix}.
\end{equation}
Diagonalizing this transfer matrix and taking the limit of a large number of domains, so that $\left(1-\tfrac{3}{2}P_0\right)^{n+1}\to\exp\left[-3P_0 z/(2z_{\rm dom})\right]$, the fluxes after propagating a distance $z$ are
\begin{equation}
I_{\gamma}(z) = I_{\gamma}(0) - P_{\gamma\rightarrow a}\left(I_{\gamma}(0)-2I_a(0)\right),
\end{equation}
\begin{equation}
I_a(z) = I_a(0) + P_{\gamma\rightarrow a}\left(I_{\gamma}(0)-2I_a(0)\right),
\end{equation}
from where we can read off the axion to photon conversion probability as~\cite{Csaki:2001yk,Grossman:2002by} 

\begin{equation}\label{eq:multi-domain-conversion}
P_{\gamma\rightarrow a} = \frac{1}{3}\left(1-e^{-z/z_{\rm decay}}\right) \,\,\, \text{with}\,\,\,
z_{\rm decay} = \frac{2z_{\rm dom}}{3P_0}.
\end{equation}

\subsection{Multi-Domain: Variable Domain Length}
\label{subsec:multi-domain-random-both}

 In Sec.~\ref{subsec:multi-domain-random-direction} we have reviewed the axion--photon oscillation probability for multiple magnetic field domains of equal length. But in astrophysical scenarios, it is rarely the case that the domains will have equal length. 
In this section we present a simplified treatment of axion--photon conversion in a region with multiple magnetic field domains where not only the magnetic field orientation changes but also the domain lengths vary from one domain to another.

We assume that the total path length traversed by the axion/photon spans multiple magnetic domains. Within each domain, the transverse magnetic field has the same magnitude across all domains, while its orientation is randomized independently in each domain, as before. Unlike the previous section, however, we now let the domain lengths vary:  the length of the $n-$th domain $z_n$ is drawn independently from a truncated Gaussian distribution with mean $\langle z_{\rm dom} \rangle$ and standard deviation $\sigma_{\rm dom}$.
 
To calculate the conversion probability of an axion into a photon after traveling a fixed distance $z$, we start from the recursion relation in Eq.~\eqref{matrix_multiply_1}. This expresses the photon and axion intensities after traversing $n+1$ domains, $I_{\gamma}^{n+1}$ and $I_a^{n+1}$, in terms of their values after $n$ domains
\begin{equation}\label{matrix_multiply_1}
\begin{pmatrix}
I_{\gamma}^{n+1}\\
I_a^{n+1}
\end{pmatrix}
=
\begin{pmatrix}
1-\dfrac{1}{2}P_0(z_{n+1}) & P_0(z_{n+1})\\
\dfrac{1}{2}P_0(z_{n+1}) & 1-P_0(z_{n+1})
\end{pmatrix}
\begin{pmatrix}
I_{\gamma}^{n}\\
I_a^{n}
\end{pmatrix},
\end{equation}
where $z_{n+1}$ is the length of the $(n+1)$th domain and $P_0(z_{n+1})$ is the axion--photon transition probability accumulated while crossing it. Unlike the equal-length case, $P_0$ now varies from domain to domain, since it depends explicitly on the (random) length of each domain.

Applying the same recursion once more to express $I_{\gamma}^{n}, I_a^{n}$ in terms of $I_{\gamma}^{n-1}, I_a^{n-1}$, we obtain
\begin{equation}
\begin{aligned}
\begin{pmatrix}
I_{\gamma}^{n+1}\\
I_a^{n+1}
\end{pmatrix}=
\begin{pmatrix}
1-\dfrac{1}{2}P_0(z_{n+1}) & P_0(z_{n+1})\\
\dfrac{1}{2}P_0(z_{n+1}) & 1-P_0(z_{n+1})
\end{pmatrix}\times \\[6pt]
\begin{pmatrix}
1-\dfrac{1}{2}P_0(z_{n}) & P_0(z_{n})\\
\dfrac{1}{2}P_0(z_{n}) & 1-P_0(z_{n})
\end{pmatrix}
\begin{pmatrix}
I_{\gamma}^{n-1}\\
I_a^{n-1}
\end{pmatrix},
\end{aligned}
\end{equation}
with the $(n+1)$th-domain matrix acting to the left of the $n$th-domain matrix, reflecting the fact that domains are traversed sequentially in order.

Iterating this procedure back to the initial intensities $I_{\gamma}(0)$ and $I_a(0)$, and keeping terms to first order in the domain probabilities, we obtain
\begin{equation}
\begin{pmatrix}
I_{\gamma}(n+1)\\
I_a(n+1)
\end{pmatrix}
=
\begin{pmatrix}
1-\dfrac{1}{2}\sum_{i=1}^{n+1} P_i & \sum_{i=1}^{n+1} P_i\\
\dfrac{1}{2}\sum_{i=1}^{n+1} P_i & 1-\sum_{i=1}^{n+1} P_i
\end{pmatrix}
\begin{pmatrix}
I_{\gamma}(0)\\
I_a(0)
\end{pmatrix},
\end{equation}
where $P_i \equiv P_0(z_i)$ is the single-domain transition probability associated with the $i$th domain, of length $z_i$. Since $P_i \ll 1$ for all $i$, cross terms of $\mathcal{O}(P_i P_j)$ and higher, generated by the product of the $n+1$ transfer matrices, are systematically neglected, leaving the leading-order result above in which only the sum $\sum_i P_i$ appears.

Let us now suppose that, over the course of the journey from source to observer, a domain of length $z_1$ recurs $n_1$ times, a domain of length $z_2$ recurs $n_2$ times, and so on, so that the total path length can be written as $z = z_1 n_1 + z_2 n_2 + \dots$, following the standard domain treatment of a randomly oriented magnetic field \cite{Kronberg1994-ly,Grasso2001}. Denoting the total number of domains traversed by $n_1+n_2+n_3+\dots = N$, the mean domain length along this path is $\langle z_{\rm dom} \rangle = z/N$; strictly, this is the sample mean of the realized domain lengths, which coincides with the population mean of the underlying truncated Gaussian distribution only in the limit $N\to\infty$, by the strong law of large numbers \citep{Feller1968}, a limit we take by the end of this paragraph. With these definitions, the sum can be recast as
\begin{equation}
\begin{aligned}
\sum_{i=1}^{n+1} P_i 
    &=\frac{z}{\langle z_{\rm dom} \rangle}\sum_{k=1}^{M}\frac{n_k}{N}P_k,
\end{aligned}
\end{equation}
where $M$ is the number of distinct domain lengths appearing along the path, and $k$ indexes them. Each factor $n_k/N$ is thus a weight, given by the fraction of domains with length $z_k$; in the limit of a large number of domains ($N\to\infty$), this weight approaches the probability density $f(z)$ of drawing a domain of length $z$ from the underlying truncated Gaussian domain-length distribution with mean $\langle z_{\rm dom}\rangle$ and standard deviation $\sigma_{\rm dom}$. Since domain lengths are taken to be positive definite, $f(z)$ is understood throughout as the truncated Gaussian probability density,
\begin{equation}
\begin{aligned}
 f(z) &= \xi^{-1}\exp\left[{-\frac{(z-\langle z_{\rm dom}\rangle)^2}{2\sigma_{\rm dom}^2}}\right], \qquad z\ge0,
 \end{aligned}
\end{equation}
with
\begin{equation}
 \xi={\sqrt{\dfrac{\pi}{2}}}\sigma_{\rm dom} \left[1 + \mathrm{Erf}\left(\dfrac{\langle z_{\rm dom}\rangle}{\sqrt{2}\sigma_{\rm dom}} \right) \right],
\end{equation}
where $\mathrm{Erf}$ is the error function. The discrete sum over realized domains is thus, in this limit, a weighted average of the conversion probability over the domain-length distribution, and may be replaced by an integral,
\begin{equation}
\begin{aligned}
\sum_{i=1}^{n+1} P_i &= \frac{z}{\langle z_{\rm dom} \rangle}\int_0^\infty P_0(z)\,f(z)\,dz \\
&\equiv \frac{z}{\langle z_{\rm dom} \rangle} \langle P_0 \rangle,
\label{eq:ensemblr_averaged_probability}
\end{aligned}
\end{equation}
where $\langle P_0 \rangle$ denotes the single domain conversion probability averaged over the ensemble of domain lengths.
This implies that for variable domain length we can simply change $P_0\rightarrow\langle P_0\rangle$. A similar kind of averaging for neutrinos has also been performed in Ref.~\cite{Ohlsson:2000mj}. Ref.~\cite{Khasai:2026zid} studies a related setup, in which the photon--ALP conversion probability is averaged over an ensemble of random magnetic field realizations. Unlike us, however, in their setup each realization is drawn from a power law distribution.

The final axion to photon conversion probability becomes
\begin{equation}\begin{aligned}\label{variable_domain_prob_eqn}
P_{\gamma\rightarrow a}
&=\frac{1}{3}
\left(1-e^{-z/\langle z_{decay}\rangle}\right),
\end{aligned}
\end{equation}
where $\langle z_{decay}\rangle=
{2\langle z_{\rm dom}\rangle}/{3\langle P_0\rangle}$.
%
Using the expression of $P_0$ from Eq.~\eqref{17} and evaluating the integral in Eq.~\eqref{eq:ensemblr_averaged_probability} one can get
\begin{equation}\label{aveprob}
\begin{aligned}
\langle P_0\rangle
&= \dfrac{(\sin 2\theta)^2}{\xi}
\int_0^\infty
\sin^2\left(\frac{\Delta_{\rm osc}}{2}z\right)
\exp\left[-\frac{(z-\langle z_{\rm dom}\rangle)^2}
{2\sigma_{\rm dom}^2}\right]dz
\\[8pt]
&=\frac{(\sin 2\theta)^2}{2}
\left[
1-\exp\left(-\frac{\Delta_{\rm osc}^2
\sigma_{\rm dom}^2}{2}\right)
\cos\left(\Delta_{\rm osc}\langle z_{\rm dom}\rangle\right)
\delta
\right],
\end{aligned}
\end{equation}
and
\begin{equation}
\delta=
\frac{(1+A)-B\tan(\Delta_{\rm osc}
\langle z_{\rm dom}\rangle)}
{1+\operatorname{Erf}\left(
\frac{\langle z_{\rm dom}\rangle}
{\sqrt{2}\sigma_{\rm dom}}
\right)},
\end{equation}
with \(A\) and \(B\) denoting the real and imaginary parts, respectively, of
\begin{equation}
A+iB=
\operatorname{Erf}\left(
\frac{\langle z_{\rm dom}\rangle
+i\Delta_{\rm osc}\sigma_{\rm dom}^2}
{\sqrt{2}\sigma_{\rm dom}}
\right).
\end{equation}
For $\Delta_{osc}\langle z_{\rm dom}\rangle\ll1$
\qquad
\begin{equation}
\begin{aligned}
    \langle P_0\rangle&=\frac{(\sin2\theta)^2}{2}\left[1-\exp\left({-\frac{\Delta_{osc}^2\sigma_{\rm dom}^2}{2}}\right)\cos \left(\Delta_{osc}\langle z_{\rm dom}\rangle\right) \delta\right]\\[6pt]
    &=\frac{4\Delta_{a\gamma}^2}{\Delta_{osc}^2}\left[\frac{\Delta_{osc}^2\langle z_{\rm dom}\rangle^2}{4}+\frac{\Delta_{osc}^2\sigma_{\rm dom}^2}{4}\right]\\[6pt]
    &=\Delta_{a\gamma}^2\langle z_{\rm dom}\rangle^2+\Delta_{a\gamma}^2\sigma_{\rm dom}^2
\end{aligned}
\end{equation}

\begin{figure}[t!]
    \centering
    \includegraphics[width=1\linewidth]{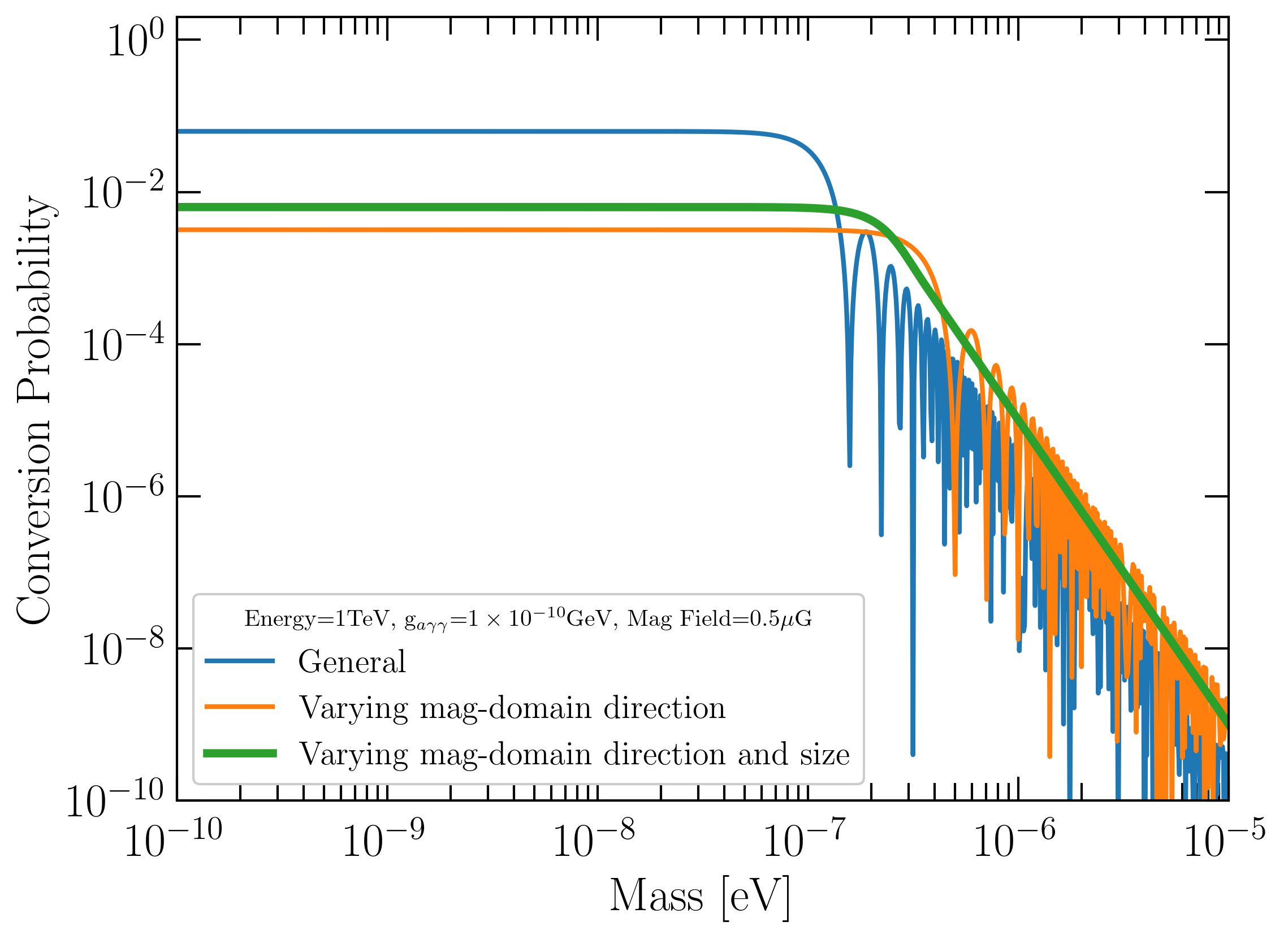}
\caption{Transition probability as a function of mass (in eV), simulated for an energy of $1 \text{ TeV}$, a coupling constant of $g_{a\gamma} = 1 \times 10^{-10} \text{ GeV}^{-1}$, and a magnetic field of $0.5 \, \mu\text{G}$. }    
\label{prob_vs_mass}
\end{figure}

Here we neglected the higher order terms and assumed $\sigma_{\rm dom}\sim \langle z_{\rm dom}\rangle$ and $\delta\rightarrow1$. We can see that $\langle P_0\rangle$ for $\Delta_{osc}\langle z_{\rm dom}\rangle\ll1$ is greater  than the usual probability by a term $\Delta_{a\gamma}^2\sigma_{\rm dom}^2$ compared to Eq. \eqref{singledomain}.

Now when $\sigma_{\rm  dom}\rightarrow 0$, which means uniform domain length, we recover the standard oscillation probability given by Eq.\eqref{17}
\begin{equation}\begin{aligned}
     \langle P_0\rangle&=\int_0^\infty P_0(z)\delta_D(z-\langle z_{\rm dom} \rangle)dz,\\[6pt]
     &=(\sin2\theta)^2
\sin^2\left(\frac{\Delta_{\rm osc}}{2}z_{\rm dom}\right),
\end{aligned}\end{equation}
As $\sigma_{\rm dom}\rightarrow0$, $f(z)\rightarrow\delta_D(z-\langle z_{\rm dom} \rangle)$, where $\delta_D$ denotes the Dirac delta function.

Fig.~\ref{prob_vs_mass} shows the axion--photon conversion probability as a function of mass for three distinct magnetic-domain configurations. The simple case (blue solid line), in which the domains are fixed in both size and orientation, exhibits a steady probability at lower masses before transitioning into rapid, deep oscillations for masses above $10^{-7}~\mathrm{eV}$. Randomizing the domain orientation alone (orange solid line) yields a slightly lower initial probability and damps the oscillatory behavior at higher masses. Randomizing both the domain orientation and size (green solid line) suppresses the oscillatory structure entirely, producing smooth, non-oscillatory behavior across the full mass range shown.

\begin{figure}[t!]
    \centering
    \includegraphics[width=1\linewidth]{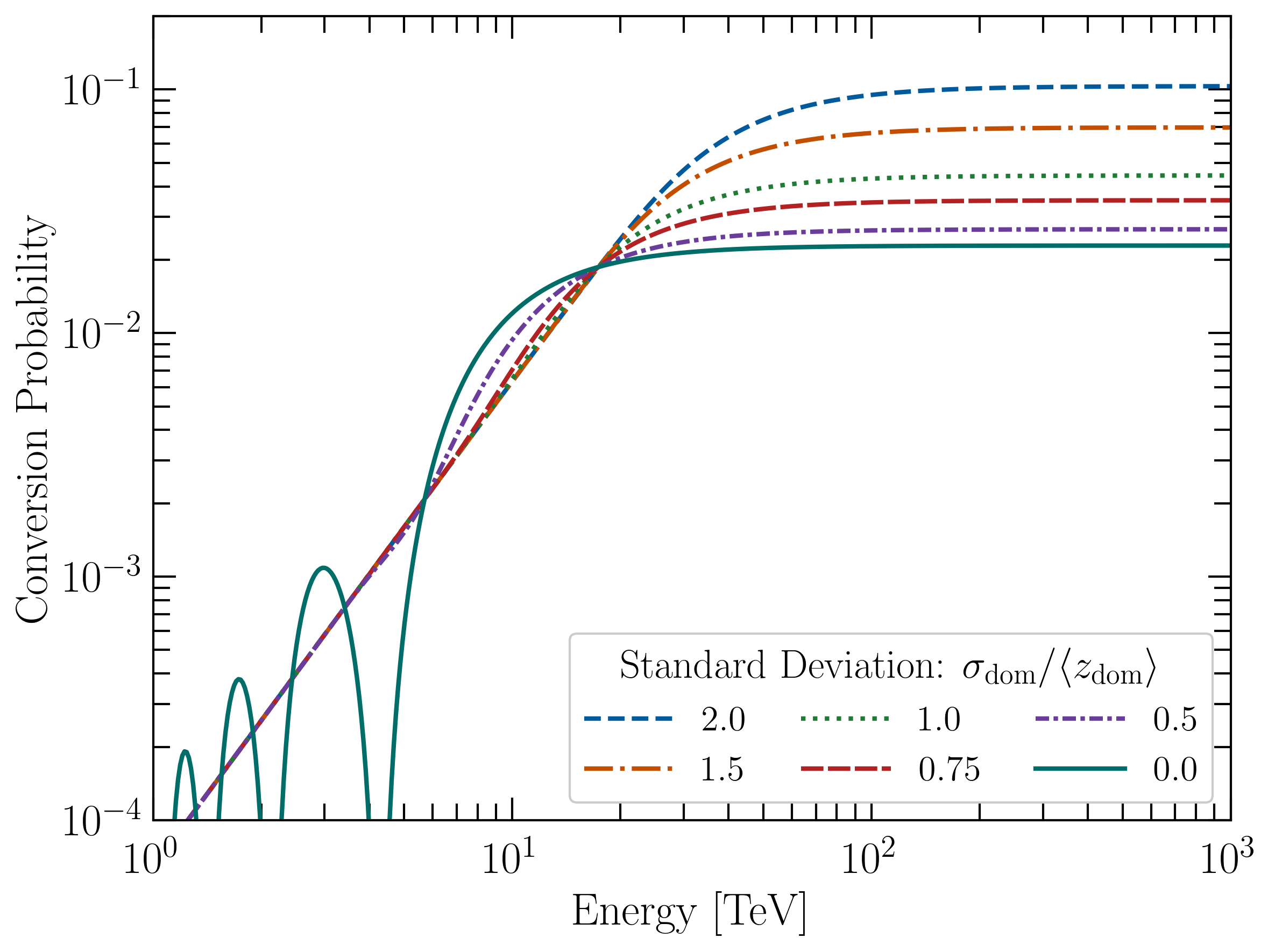}
  \caption{Energy dependence of the photon--axion conversion probability, $P_{\gamma\to a}$, for different values of the standard deviation $\sigma_{\rm dom}$ of the magnetic-domain length distribution, normalized to the mean coherence length $\langle z_{\rm dom}\rangle$. The mass $m_a=4.2\times 10^{-7}$eV, and the coupling $g_{a\gamma}=1.4\times10^{-10}$GeV$^{-1}$, these values are taken from \cite{Qin:2025olt}.}
    \label{pic_prob_diff_sigma}
\end{figure}

A complementary comparison of the conversion probability as a function of energy is presented in Fig.~\ref{pic_prob_diff_sigma} for several values of $\sigma_{\rm dom}/\langle z_{\rm dom}\rangle$. The solid green curve, corresponding to $\sigma_{\rm dom}/\langle z_{\rm dom}\rangle = 0$, recovers the fixed-domain case. For nonzero $\sigma_{\rm dom}/\langle z_{\rm dom}\rangle$, the conversion probability is clearly enhanced relative to this baseline, demonstrating that domain-length fluctuations systematically boost the axion--photon conversion probability. 

\section{Ultra High Energy Photons from GRB~221009A}
\label{sec:grb}

\begin{figure}[t!]
    \centering
    \includegraphics[width=1\linewidth]{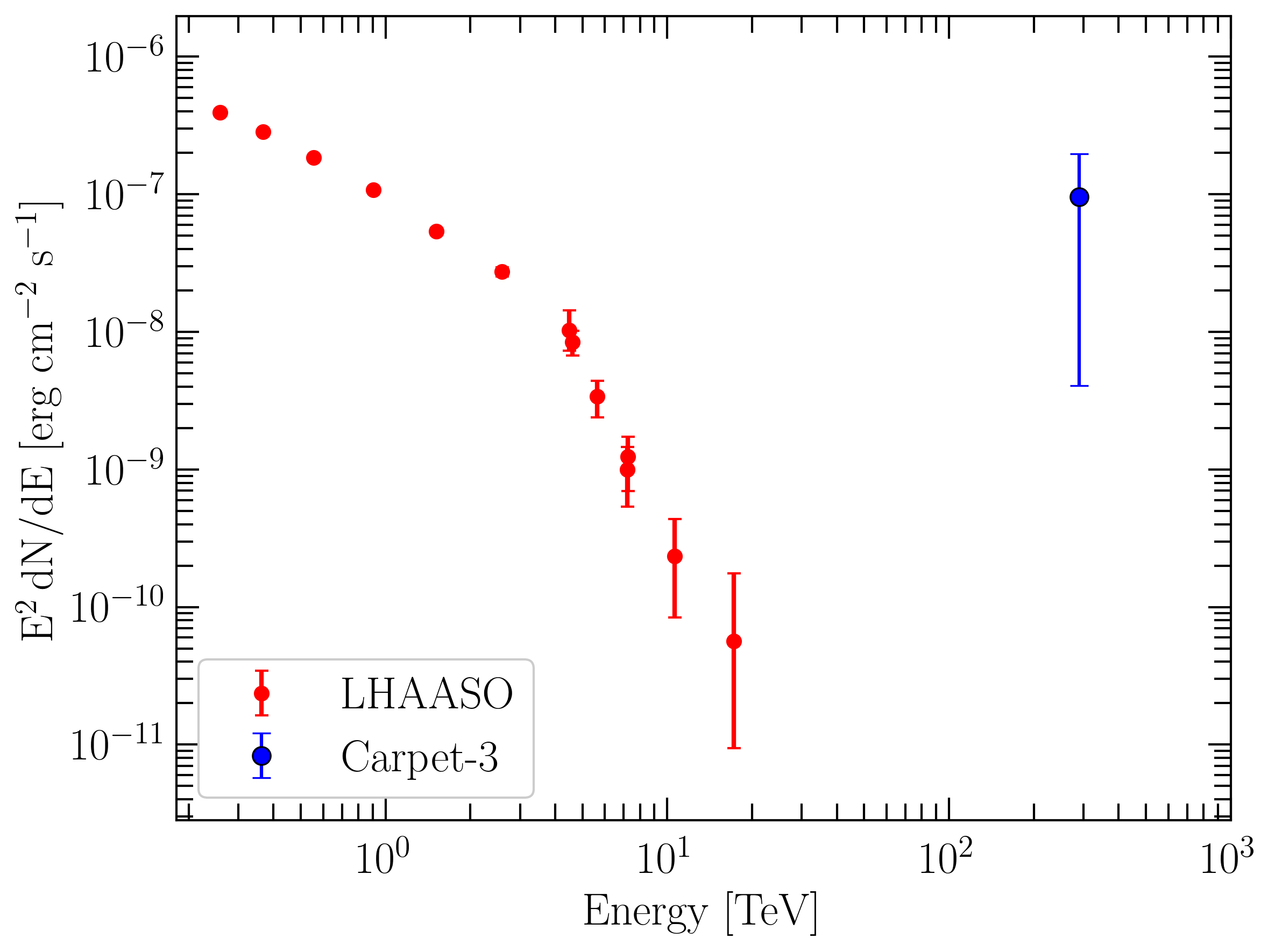}
    \caption{Observed spectral energy distribution, E$^2\frac{dN}{dE}$ of GRB~221009A as observed by LHAASO and Carpet-3.}
    \label{data_figure}
\end{figure}

GRB~221009A is an extremely bright gamma-ray burst~\cite{Burns:2023oxn} associated with a Type Ic supernova~\cite{Blanchard:2023zls} located in the constellation Sagitta at redshift $\hat z_s = 0.151$~\cite{2022GCN.32648....1D,2022GCN.32686....1C}. The LHAASO observatory reported a huge flux of very high energy and ultra-high energy (UHE) photons from this GRB, with the highest recorded photon energy being $18$~TeV~\cite{LHAASO:2023kyg,LHAASA:2023pay}. Additionally, the Carpet-3 observatory reported the observation of a UHE photon with energy $\sim 300$~TeV~\cite{Carpet3:2022uhy,Carpet-3Group:2025fcs}. In Fig.~\ref{data_figure} we show the reconstructed photon flux as reported by these experiments.

However, the observation of these UHE photons from GRB~221009A presents a direct challenge to our understanding~\cite{Galanti:2022chk}. UHE photons emitted from distant astrophysical sources such as GRB~221009A travel enormous distances through the intergalactic medium (IGM) and reach the Earth. The IGM is permeated by a background of low-energy photons constituting the Cosmic Microwave Background (CMB)~\cite{1965ApJ...142..419P}, the Extragalactic Background Light (EBL)~\cite{Hauser:2001xs,Dwek:2012nb} and the Cosmic Radio Background~\cite{Protheroe:1996si,Fixsen:2009xn}. While traveling through the IGM, UHE photons can interact with these low-energy background photons to produce electron-positron pairs via pair production ($\gamma\gamma \rightarrow e^+e^-$)~\cite{Nikishov:1962rmq,1967PhRv..155.1404G}, which can lead to their extinction and a significant loss of flux over distances spanning several megaparsecs~\cite{Stecker:1995tm}.

To understand why this is happening, let us consider that UHE photons are produced at some distant extragalactic source at redshift $\hat z_s$. The observed intensity of photons at energy $E_\gamma$~\cite{Stecker:1995tm,Dwek:2012nb} at Earth is given by
\begin{equation}
    I_{\rm obs}(E_\gamma) = I_0(E_0)e^{-\tau_\gamma(E_\gamma)},
\end{equation}
where $I_0$ is the initial intensity, $E_0 = E_\gamma(1+\hat z_s)$ is the initial energy which is redshifted down to $E_\gamma$. The factor $e^{-\tau_\gamma(E_\gamma)}$ is the photon survival probability with $\tau_\gamma(E_\gamma)$ being the optical depth (opacity) of the IGM to photons of observed energy $E_\gamma$ emitted by a source at redshift $\hat z_s$. It accounts for the cumulative probability of pair production along the line of sight, and is given by
\begin{equation}
    \tau_\gamma(E_\gamma) = \int_0^{\hat z_s} d\hat z\, \frac{dl}{d\hat z} \int_{-1}^{1} d\mu\, \frac{1-\mu}{2} \int_{\epsilon_{\rm th}}^{\infty} d\epsilon\; n_\gamma(\epsilon, \hat z)\, \sigma_{\gamma\gamma}(s),
    \label{eq:tau}
\end{equation}
where $\epsilon$ is the energy of the background photon at redshift $\hat z$, $\mu=\cos\theta$ is the cosine of the angle between the momenta of the two interacting photons and $n_\gamma(\epsilon,\hat z)$ is the differential number density of the background photons, which can be obtained from~\cite{Dominguez:2010bv}. The line element in a flat $\Lambda$CDM cosmology is
\begin{equation}
    \frac{dl}{d\hat z} = \frac{1}{(1+\hat z)H(\hat z)}, \quad H(\hat z)=H_0\sqrt{\Omega_m(1+\hat z)^3+\Omega_\Lambda}.
\end{equation}
where $H_0$ is the Hubble constant, $\Omega_m$ is the matter density parameter and $\Omega_\Lambda$ is the dark-energy density parameter, with values taken from Ref.~\cite{Planck:2018vyg}.
The interaction is governed by the Breit-Wheeler cross-section~\cite{Breit:1934zz,1967PhRv..155.1404G},
\begin{align}
    \sigma_{\gamma\gamma}(s) &= \frac{3\sigma_T}{16}(1-\beta^2)\left[(3-\beta^4)\ln\frac{1+\beta}{1-\beta} - 2\beta(2-\beta^2)\right], \nonumber\\
    \beta &= \sqrt{1-\frac{4m_e^2}{s}},
\end{align}
with $\sigma_T$ the Thomson cross-section and $s$ being the center of mass energy. 

Now at a fixed redshift $\hat z$ where the photon has energy $E=E_\gamma(1+\hat z)$, pair production is kinematically allowed only for $s\geq 4m_e^2$. This fixes the threshold energy of the background photons that can pair produce electrons to
\begin{equation}
    \epsilon_{\rm th} = \frac{2m_e^2}{E(1-\mu)}.
    \label{eq:threshold_energy}
\end{equation}
From Eq.~\eqref{eq:threshold_energy} one can see that a photon of energy $E$ is absorbed most efficiently by background photons with $\epsilon \sim 2m_e^2/E\simeq 0.5\,(1~{\rm TeV}/E)$ eV. Thus depending on the incoming UHE photon energy, different backgrounds dominate in different energy windows. The EBL~\cite{Hauser:2001xs,Dwek:2012nb} with energies of the order of few to tens of eV is responsible for the attenuation of $E_\gamma\sim 0.1$-$100$ TeV photons, whereas for $E_\gamma\gtrsim 100$ TeV the relevant target photons are the CMB photons with energies of the order of $10^{-4}$ eV. At even higher energies ($E\gtrsim 10^{18}$ eV), the cosmic radio background~\cite{Protheroe:1996si} becomes the dominant target. 

\begin{figure}[t!]
    \centering
    \includegraphics[width=1\linewidth]{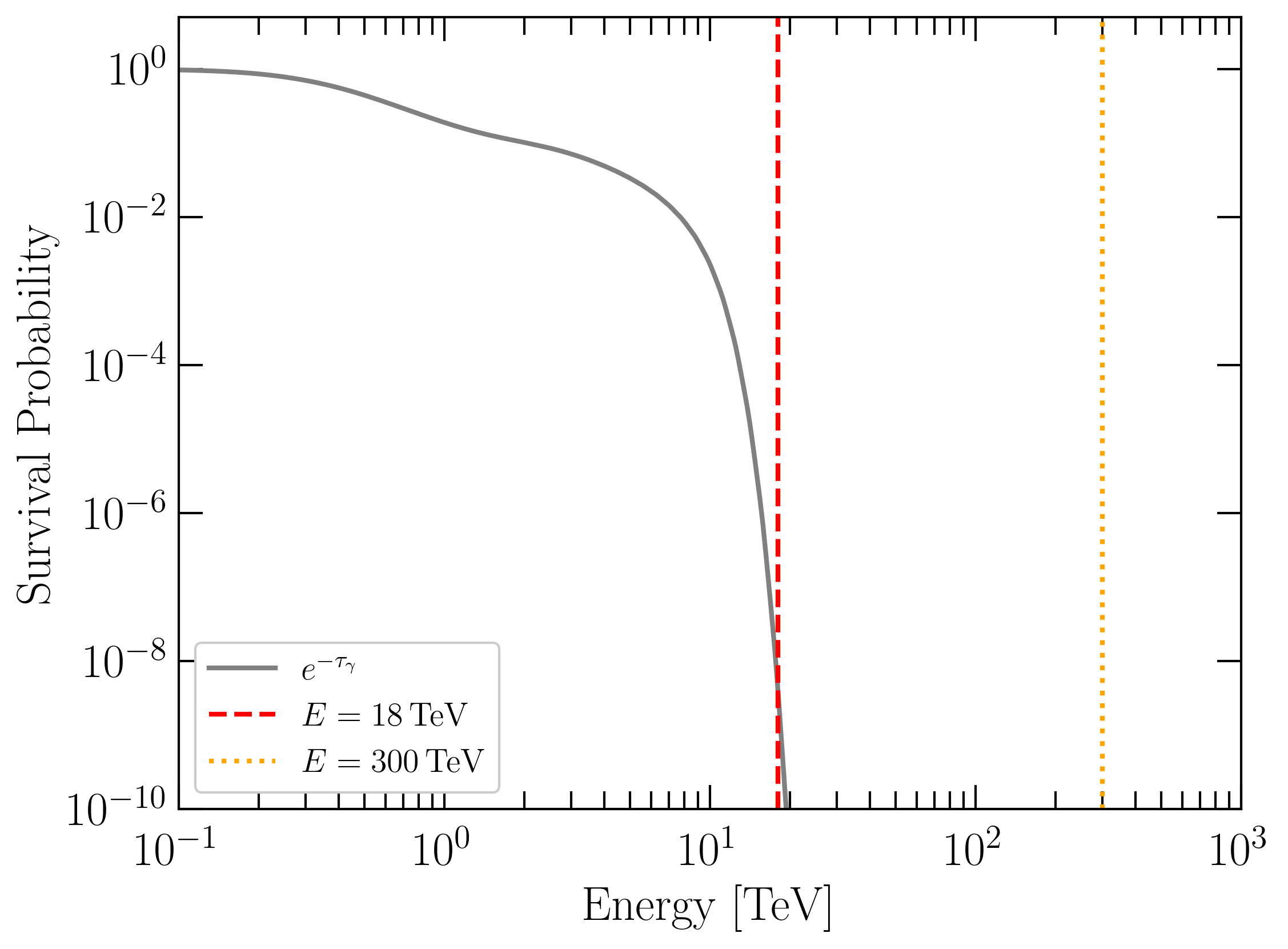}
    \caption{Photon survival probability \(P_{\gamma\rightarrow\gamma}^{I}\) during propagation through the intergalactic medium accounting for photon attenuation. At \(E=18\,\mathrm{TeV}\), the survival probability is of order \(10^{-9}\), while at \(E=300\,\mathrm{TeV}\) it becomes \(\ll10^{-10}\). Thus, the survival probability becomes extremely small at high energies and is insufficient to account for a significant flux of ultra-high-energy photons reaching Earth. The vertical dashed lines indicate \(E=18\,\mathrm{TeV}\) and \(E=300\,\mathrm{TeV}\).}
    \label{attenuation_figure}
\end{figure}

Fig.~\ref{attenuation_figure} shows the survival probability $P^{\rm I}_{\gamma \to \gamma} = e^{-\tau_\gamma}$ as a function of the observed photon energy for a source at $\hat z_s=0.151$. One can clearly see that the Universe is transparent for $E_\gamma \lesssim 1$ TeV, while $e^{-\tau_\gamma}$ falls steeply above $\sim 10$ TeV. At $18$ TeV (red dashed line), $\tau_\gamma = 19.47$, corresponding to a suppression of almost nine orders of magnitude. The curve drops below the plotted range by $\sim 20$ TeV and never reaches the $300$ TeV line (orange dashed), so a photon at that energy cannot be explained by standard propagation.

\section{Propagation of UHE Photons in the Presence of Axions}
\label{sec:photon_prop}
To explain the observation of UHE photons from GRB~221009A one needs physics beyond the Standard Model, and axions present themselves as a natural candidate~\cite{Galanti:2022chk,Qin:2025olt,Satunin:2025hbk}. Host galaxies of GRBs, just like our Milky Way, are expected to host large scale magnetic fields, in which photons can convert into axions. Being weakly interacting, the axions propagate through the IGM unimpeded, since they cannot pair-produce on the EBL/CMB, and can reconvert into photons in the Galactic magnetic field.

Including photon--ALP conversion in the source (S), intergalactic (I) and Milky Way (M) regions, the total photon survival probability is modified to
\begin{equation}
\begin{aligned}\label{total_survival_of_photon}
P_{\gamma\rightarrow\gamma} &= P_{\gamma\rightarrow\gamma}^{\text{S}} \left( P_{\gamma\rightarrow\gamma}^{\text{I}} P_{\gamma\rightarrow\gamma}^{\text{M}} + P_{\gamma\rightarrow a}^{\text{I}} P_{a\rightarrow\gamma}^{\text{M}} \right)\\[6pt]
& \quad + P_{\gamma\rightarrow a}^{\text{S}} \left( P_{a\rightarrow\gamma}^{\text{I}} P_{\gamma\rightarrow\gamma}^{\text{M}} + P_{a\rightarrow a}^{\text{I}} P_{a\rightarrow\gamma}^{\text{M}} \right),
\end{aligned}
\end{equation}
where $P^{X}_{i\to j}$ is the probability that a state $i$ entering region $X$ exits as state $j$. Each term is a product along one of the possible paths $\mathrm{S}\to\mathrm{I}\to\mathrm{M}$. For example, $P^{\rm S}_{\gamma\to a}P^{\rm I}_{a\to a}P^{\rm M}_{a\to\gamma}$ describes a photon converting to an ALP in the source, crossing the IGM as an ALP, and reconverting to a photon in the Milky Way. Since these environments host strong large-scale magnetic fields, the survival probability, and hence the observed UHE photon flux, can be modified significantly. In this section we show that this effect can help explain the UHE photons from GRB~221009A observed by LHAASO and Carpet-3.

To do so, we model the intrinsic, unattenuated photon spectrum of GRB~221009A as a simple unbroken power law~\cite{Qin:2025olt},
\begin{equation}
\label{eq:power_law}
\frac{dN_{\rm int}}{dE}
=\Phi_0
\left(\dfrac{E}{E_0}\right)^{-\Gamma},
\end{equation}
where $\Phi_0$ is the normalization, $E_0=1$ TeV, and $\Gamma$ is the spectral index. The observed spectrum, modified by photon-ALP conversion and reconversion as well as absorption on background photons, is
\begin{equation}
\frac{dN_{\rm obs}}{dE}
=
\frac{dN_{\rm int}}{dE}\,
P_{\gamma\rightarrow\gamma}(E;m_a,g_{a\gamma}),
\end{equation}
where $P_{\gamma\rightarrow\gamma}$ is the total survival probability of Eq.~\eqref{total_survival_of_photon}. The corresponding observed spectral energy distribution is
\begin{equation}
E^2\frac{dN_{\rm obs}}{dE}
=
E^2\frac{dN_{\rm int}}{dE}\,
P_{\gamma\rightarrow\gamma}(E;m_a,g_{a\gamma}).
\label{eq:flux}
\end{equation}
This is the flux of UHE photons expected to reach Earth, including the effects of photon--ALP conversion.

\begin{figure}[t!]
    \centering
    \includegraphics[width=1\linewidth]{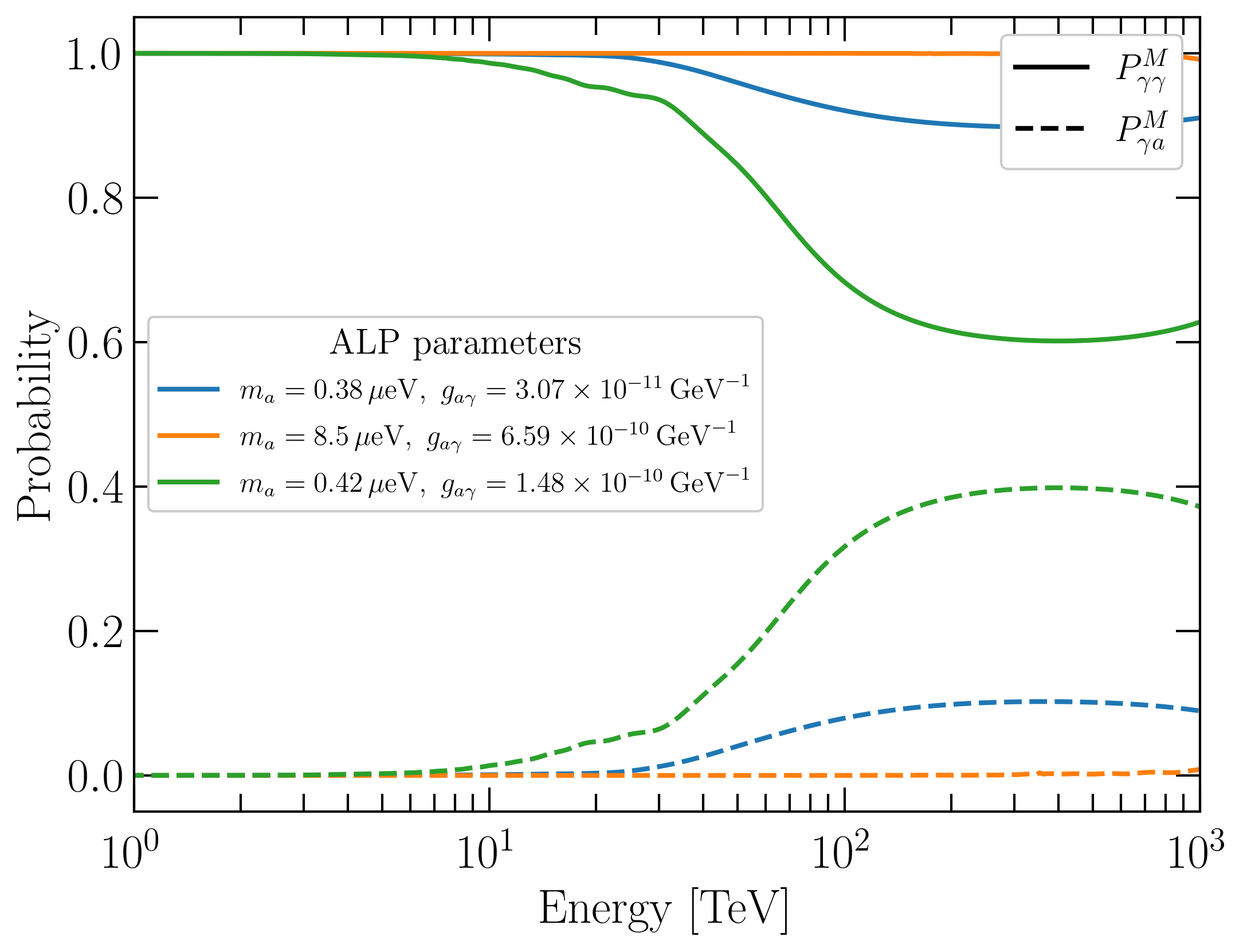}
    \caption{Energy dependence of the photon survival probability, $P_{\gamma\rightarrow\gamma}^\text{M}$ (solid lines), and the photon--ALP conversion probability, $P_{\gamma\rightarrow a}^\text{M}$ (dashed lines), for photon propagation through the Galactic magnetic field. Colors denote representative ALP parameter sets $(m_a,g_{a\gamma})$ as indicated in the legend.}
    \label{pic_prob_mw_diff_param}
\end{figure}

To model photon--ALP oscillations in the Milky Way, we use the \texttt{gammaALPs}~\cite{Meyer:202199} Python package, which builds the Galactic magnetic field from the Jansson \& Farrar model~\cite{Jansson:2012pc}. The photon survival probability $P^{\rm M}_{\gamma \rightarrow \gamma}$ and the conversion probability $P^{\rm M}_{\gamma \rightarrow a}$ for different values of the ALP mass $m_a$ and coupling $g_{a\gamma}$ are shown in Fig.~\ref{pic_prob_mw_diff_param}. Photon--ALP mixing becomes efficient at energies above $\sim10$--$100\,\mathrm{TeV}$, reducing $P^{\rm M}_{\gamma \rightarrow \gamma}$ and enhancing $P^{\rm M}_{\gamma \rightarrow a}$ correspondingly. The size of the effect depends strongly on $m_a$ and $g_{a\gamma}$.

We also use \texttt{gammaALPs} to compute the photon survival probability $P^{\rm I}_{\gamma \rightarrow \gamma}$ and the conversion probability $P^{\rm I}_{\gamma \rightarrow a}$ in the IGM. The total propagation distance of 745~Mpc is modeled as a sequence of magnetic domains of fixed length 10~Mpc, each with field strength $1$~nG~\cite{Qin:2025olt,Han:2026wdb} and a random orientation. \texttt{gammaALPs} includes the attenuation of UHE photons by EBL scattering through the factor $e^{-\tau_\gamma}$ shown in Fig.~\ref{attenuation_figure}. We have checked that photon--ALP conversion in the IGM has a negligible effect on the final photon flux, but we retain it in our treatment.

Finally, we model the magnetic field of the host galaxy (S) as a collection of $N=5$ domains within a radius of $30$~kpc, each with a fixed field strength $B=1\,\mu$G and a random orientation. The domain lengths $z_{\rm dom}$ are allowed to vary, with mean $\langle z_{\rm dom}\rangle = 6$~kpc and relative dispersion $\sigma_{\rm dom}/\langle z_{\rm dom}\rangle$. The photon--ALP conversion probability in the source is obtained from the averaged oscillation probability of Eq.~\eqref{aveprob}. For $\sigma_{\rm dom}=0$ all domains have the same length, which reproduces the fixed-domain-length scenario studied in Refs.~\cite{Qin:2025olt,PhysRevD.108.023002}. The variable-domain-length formalism is therefore a generalization of that setup.

\begin{figure}[t!]
    \centering
    \includegraphics[width=1\linewidth]{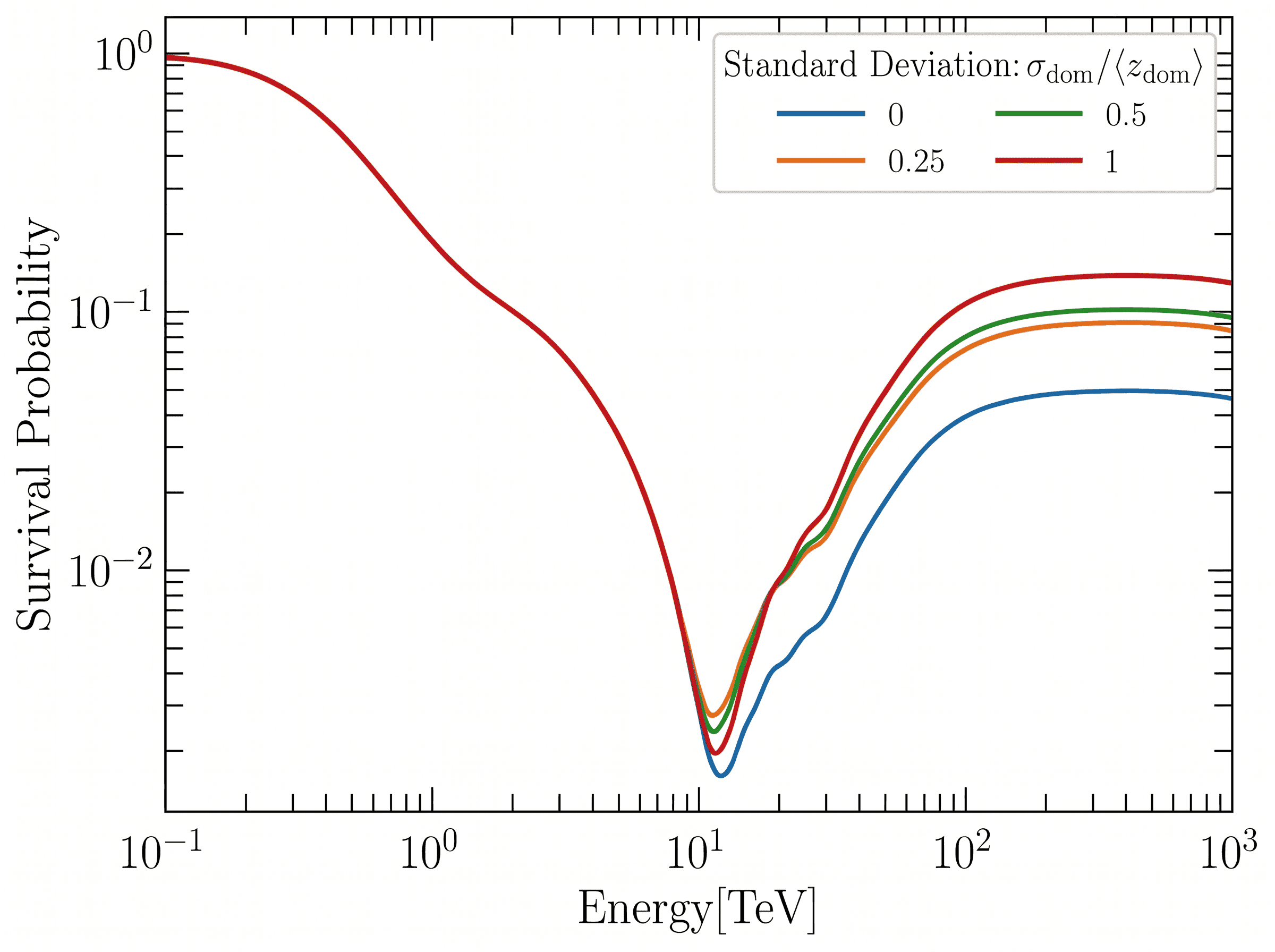}
    \caption{Energy dependence of the photon survival probability, $P_{\gamma\rightarrow\gamma}$, for a Gaussian distribution of magnetic-domain lengths characterized by a mean coherence length $\langle z_{\rm dom}\rangle$ and standard deviation $\sigma_{\rm dom}$. The legend indicates the normalized dispersion parameter $\sigma_{\rm dom}/\langle z_{\rm dom}\rangle$. Increasing the spread in domain lengths modifies the oscillation pattern and enhances the high-energy survival probability relative to the uniform-domain case ($\sigma_{\rm dom}=0$).}
    \label{pic_photon_surv_prob_grb}
\end{figure}

Fig.~\ref{pic_photon_surv_prob_grb} shows the total photon survival probability from GRB~221009A as a function of the photon energy. We see that photon--ALP conversion lifts the survival probability at high energies, and that a broader domain-length distribution in the source enhances this rise. As mentioned before, earlier studies have also considered axions as an explanation of the UHE photons from this burst, but with fixed domain lengths in the source they could not accommodate the Carpet-3 data~\cite{Qin:2025olt}, and had to invoke LIV as an additional ingredient. Allowing the domain lengths to vary raises the survival probability at the highest energies, so that the data, including Carpet-3, can be explained by photon--ALP conversion alone, without invoking LIV or any other new effect.

To quantify this, we fit the spectrum of Eq.~\eqref{eq:flux} to the LHAASO~\cite{LHAASA:2023pay,LHAASO:2023tbl} and Carpet-3~\cite{Carpet3:2022uhy,Carpet-3Group:2025fcs} data of GRB~221009A, and extract the best-fit ALP parameters together with the $1\sigma$ and $2\sigma$ viable regions in the $(m_a,g_{a\gamma})$ plane. We perform the fit for two cases: $\sigma_{\rm dom}/\langle z_{\rm dom}\rangle=0$, which reproduces the fixed-domain-length benchmark of Ref.~\cite{Qin:2025olt}, and $\sigma_{\rm dom}/\langle z_{\rm dom}\rangle=1$, the maximal domain-length dispersion we consider. The intermediate values $0.25$ and $0.5$ in Fig.~\ref{pic_photon_surv_prob_grb} are shown for illustration only and are not fitted. In both cases $P_{\gamma\rightarrow\gamma}$ is computed along S$\rightarrow$I$\rightarrow$M as described above, so the comparison isolates the effect of the domain-length distribution on the viable region.

We use the Gaussian $\chi^2$
\begin{equation}
\begin{aligned}
&\chi^2(m_a,g_{a\gamma},\Phi_0,\Gamma)
\\
&=\sum_{i=1}^{N} 
\frac{
\left[
y_i-E_i^2\dfrac{dN_{\rm obs}}{dE}
(E_i;m_a,g_{a\gamma},\Phi_0,\Gamma)
\right]
}{\sigma_i^2}^2.
\end{aligned}
\end{equation}
where $y_i$ is the measured $E^2\,dN/dE$ in the $i$-th energy bin and $\sigma_i$ its uncertainty. For asymmetric uncertainties, we conservatively take $\sigma_i$ to be the larger of the two. The source spectral parameters $(\Phi_0,\Gamma)$ of Eq.~\eqref{eq:power_law} are nuisance parameters, which we profile out by minimizing the $\chi^2$ over them at each $(m_a,g_{a\gamma})$. The best-fit ALP parameters minimize the resulting profiled $\chi^2$, with minimum $\chi^2_{\min}$, and the goodness of fit is quantified by $\chi^2_\nu=\chi^2_{\min}/\nu$, where $\nu$ is the number of data points minus the number of fitted parameters. The $1\sigma$ and $2\sigma$ regions are defined by $\Delta\chi^2\equiv\chi^2-\chi^2_{\min}<2.30$ and $6.18$ respectively~\cite{Cowan:2010js,ParticleDataGroup:2026mpi}.

\section{Results and Discussion}
\label{sec:results}

\begin{figure}[t!]
    \centering
    \includegraphics[width=1\linewidth]{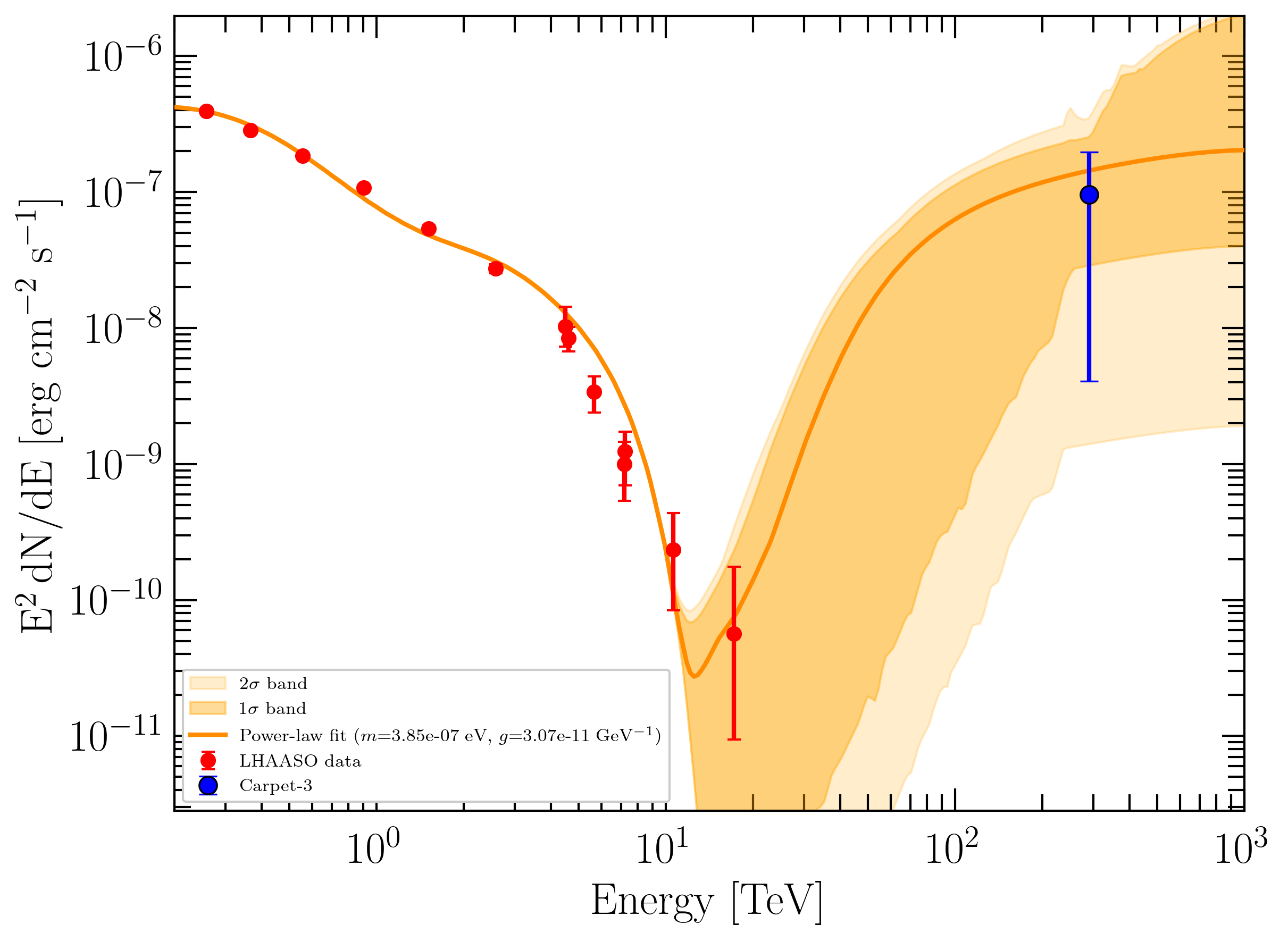}
    \includegraphics[width=1\linewidth]{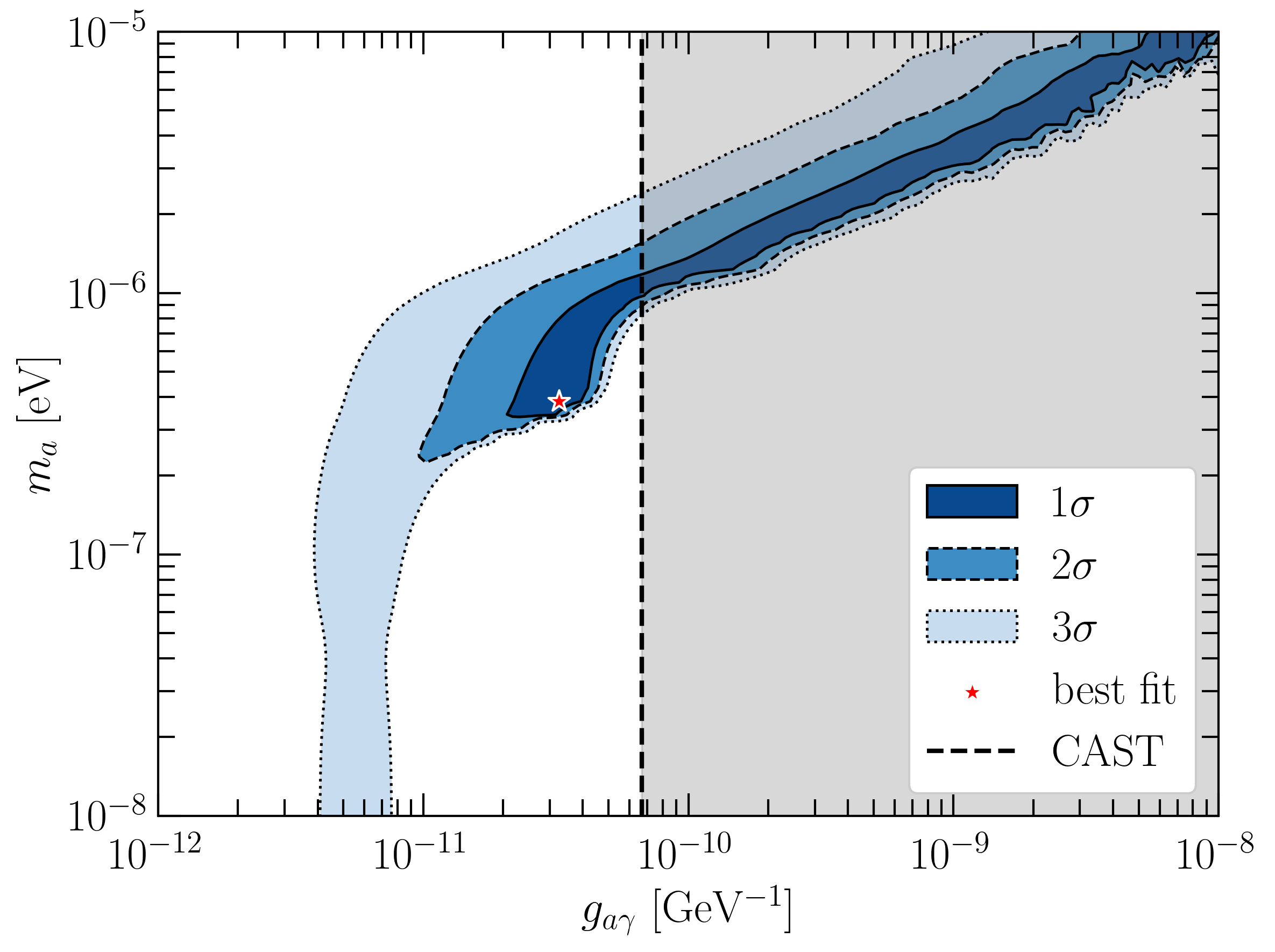}
    \caption{\textit{\bf Top panel:} Best-fit spectrum for the variable-domain-length formalism, with spectral index $\Gamma=1.63$. \textit{\bf Bottom panel:} Confidence-level contours in the $(m_a,g_{a\gamma})$ plane for the variable-domain-length formalism with a single power-law spectrum. The CAST exclusion region is also shown.}
    \label{fig:unbroken_best}
\end{figure}

\begin{figure}[t!]
    \centering
    \includegraphics[width=1\linewidth]{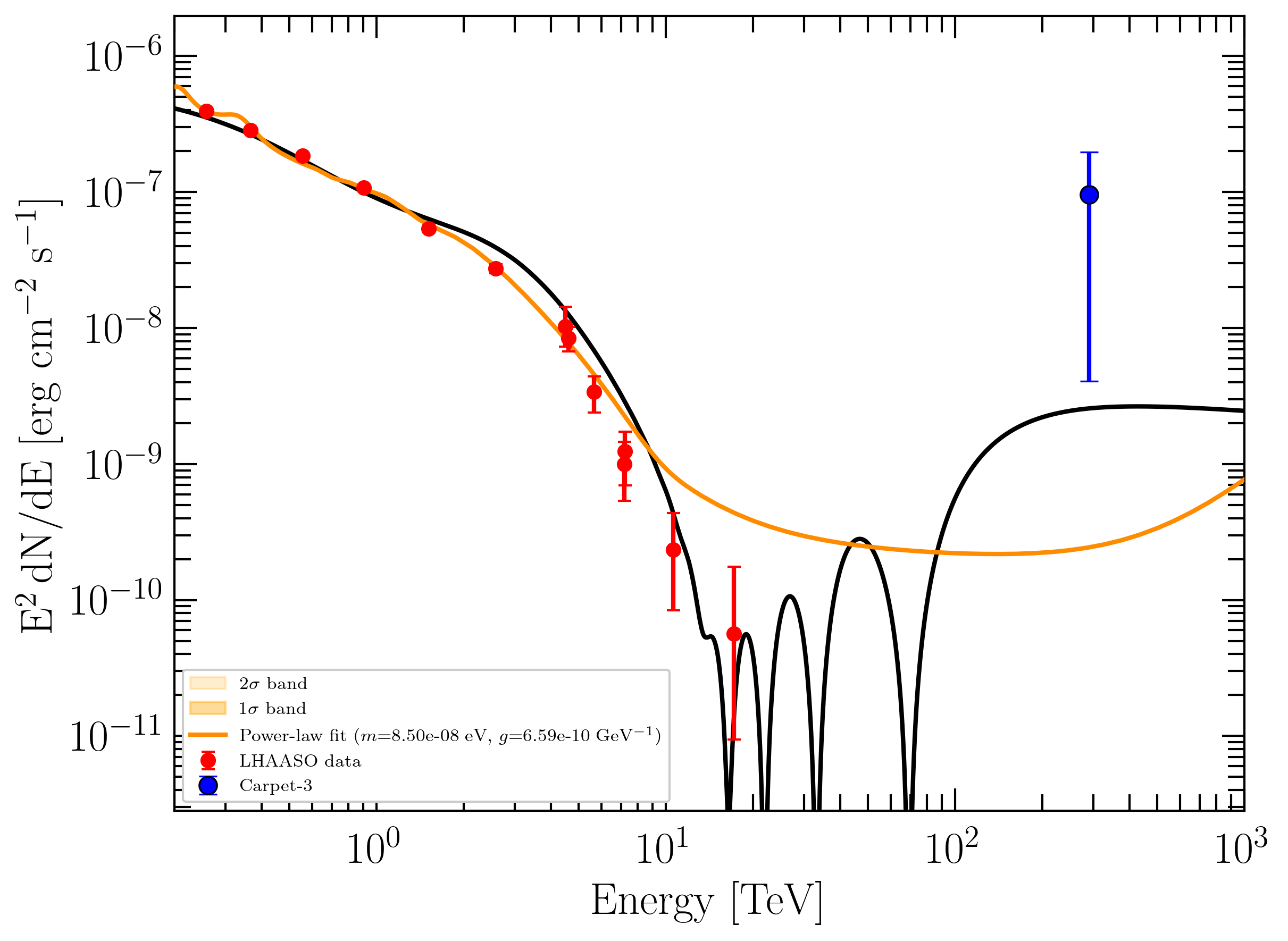}
    \includegraphics[width=1\linewidth]{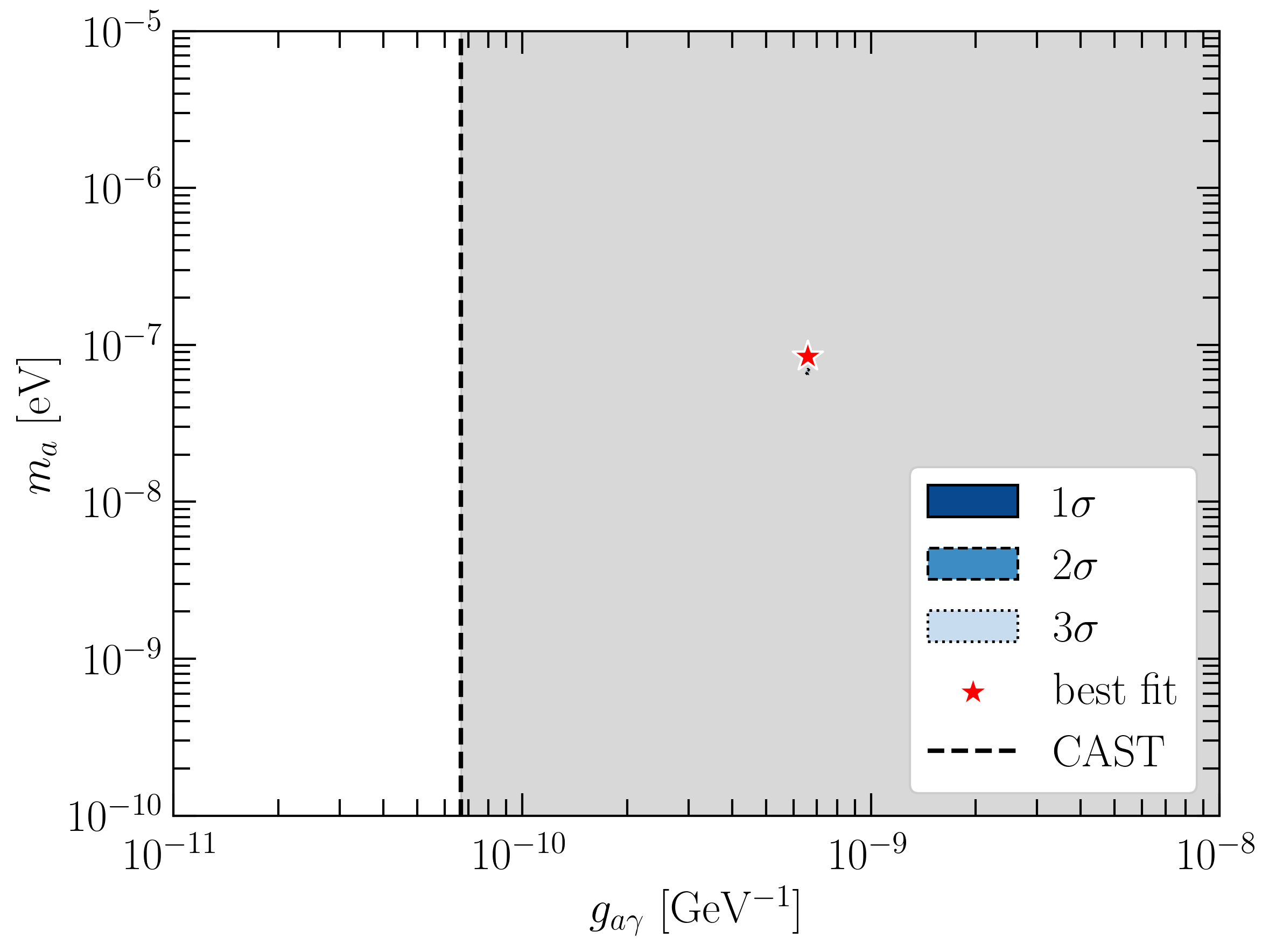}
    \caption{\textit{\bf Top panel:} Best-fit spectrum for the fixed-domain-length formalism, with spectral index $\Gamma=2.11$. The black line is the ALP-only best fit of Ref.~\cite{Qin:2025olt}. \textit{\bf Bottom panel:} Confidence-level contours in the $(m_a,g_{a\gamma})$ plane for the fixed-domain-length formalism with a single power-law spectrum. The CAST exclusion region is also shown.}
    \label{fig:fixeddomain}
\end{figure}

We scan the ALP parameter space $10^{-10}\,\mathrm{eV}\le m_a\le 10^{-5}\,\mathrm{eV}$ and $10^{-13}\,\mathrm{GeV}^{-1}\le g_{a\gamma}\le 10^{-8}\,\mathrm{GeV}^{-1}$, evaluating the spectrum and the profiled $\chi^2$ at each point. Figs.~\ref{fig:unbroken_best} and~\ref{fig:fixeddomain} show the best-fit spectra and the confidence regions in the $(m_a,g_{a\gamma})$ plane for $\sigma_{\rm dom}/\langle z_{\rm dom}\rangle=1$ (variable domain length) and $\sigma_{\rm dom}=0$ (fixed domain length), respectively. The fixed-domain fit is poor, mainly because it cannot reach the Carpet-3 point, whereas the variable-domain fit describes both the LHAASO and Carpet-3 data.

For variable domain lengths, the minimum $\chi^2$ is obtained at
\begin{equation}
    m_a = 3.85\times10^{-7}\,\mathrm{eV}, \quad g_{a\gamma}=3.07\times10^{-11}\,\mathrm{GeV}^{-1},
\end{equation}
with a fitted spectral index $\Gamma=1.63$, i.e.\ $dN_{\rm int}/dE=\Phi_0\,(E/E_0)^{-1.63}$. The best-fit spectrum is shown in the top panel of Fig.~\ref{fig:unbroken_best}, and the $1\sigma$, $2\sigma$ and $3\sigma$ regions in the bottom panel.

For fixed domain lengths, the best fit $\chi^2$ is at
\begin{equation}
    m_a = 8.50\times10^{-8}\,\mathrm{eV}, \quad g_{a\gamma}=6.59\times10^{-10}\,\mathrm{GeV}^{-1},
\end{equation}
with $\Gamma=2.11$, i.e.\ $dN_{\rm int}/dE=\Phi_0\,(E/E_0)^{-2.11}$ (Fig.~\ref{fig:fixeddomain}). This spectrum follows the LHAASO points but falls far below the Carpet-3 point at 300~TeV.

Fig.~\ref{fig:fixeddomain} also shows the best fit of Ref.~\cite{Qin:2025olt}, $m_a=4.191\times10^{-7}$~eV and $g_{a\gamma}=1.489\times10^{-10}\,\mathrm{GeV}^{-1}$, obtained for photon--ALP conversion only. Our analysis differs from theirs in the modeling of all three propagation regions. For the host galaxy, we use a $30$~kpc region with $B=1\,\mu$G divided into 5 domains of random orientation and varying length, whereas they use $10$~kpc and $0.5\,\mu$G without specifying the domain size. For the IGM, the field is weak enough that its effect is negligible, as also noted in their work. For the Milky Way, we construct the field with the Jansson \& Farrar model in \texttt{gammaALPs} along the line of sight to GRB~221009A, whereas they approximate it as $3\,\mu$G over $\sim30$~kpc and take $P^{\rm M}_{\gamma\rightarrow\gamma}\approx1$, i.e.\ they neglect conversion in the Milky Way. As shown in Fig.~\ref{pic_prob_mw_diff_param}, however, this conversion is not negligible at high energies, so it must be included when computing the flux at Earth.

We compare the viable regions with the CAST bound~\cite{Anastassopoulos:2017ftl}, $g_{a\gamma}\le6.67\times10^{-11}\,\mathrm{GeV}^{-1}$, which is independent of astrophysical modelling and holds for ALP masses up to $m_a\sim0.02$~eV, far above the masses relevant here. For fixed domain lengths, the best-fit spectrum falls below the Carpet-3 point (Fig.~\ref{fig:fixeddomain}, top panel), and the best-fit coupling, $g_{a\gamma}=6.59\times10^{-10}\,\mathrm{GeV}^{-1}$, is an order of magnitude above the CAST bound (Fig.~\ref{fig:fixeddomain}, bottom panel). The confidence regions around it are too small to be visible in the figure, but they lie entirely within the CAST-excluded region. The fixed-domain model therefore cannot accommodate the Carpet-3 data without conflicting with laboratory limits.

The situation is different for variable domain lengths. The best-fit spectrum reaches the Carpet-3 point while following the LHAASO data (Fig.~\ref{fig:unbroken_best}, top panel). The best-fit coupling, $g_{a\gamma}=3.07\times10^{-11}\,\mathrm{GeV}^{-1}$, lies a factor of about two below the CAST bound, and the $1\sigma$, $2\sigma$ and $3\sigma$ regions all extend across the CAST line (Fig.~\ref{fig:unbroken_best}, bottom panel). Part of each region therefore lies at couplings allowed by CAST, so the variable-domain model is not excluded. The portions above the CAST line are excluded, and the viable parameter space is the intersection of the confidence regions with the CAST-allowed region. Compatibility with laboratory bounds thus cannot be judged from the best-fit point alone, but requires the full confidence region.

Our results also bear on the earlier literature. The ALP-only best fit of Ref.~\cite{Qin:2025olt} with fixed magnetic domains, $g_{a\gamma}=1.489\times10^{-10}\,\mathrm{GeV}^{-1}$, lies a factor of about two above the CAST bound, and fails to reproduce the Carpet-3 point (Fig.~\ref{fig:fixeddomain}, top panel) requiring the introduction of LIV as an additional ingredient. Our fixed-domain analysis, which includes the Milky Way conversion neglected there, reaches the same conclusion: a single domain length cannot simultaneously fit the data and respect the CAST bound. Allowing a variability of the domain lengths of the source removes this tension, so that the Carpet-3 data can be explained by photon--ALP conversion alone, without LIV or any other physics beyond the axion.

Thus our results highlight the fact that the treatment of the magnetic-domain structure strongly affects the conversion probability, the spectral fit and the inferred ALP parameter space.

\section*{Acknowledgments}

The authors sincerely acknowledge Satyaki Bhattacharya for the valuable discussions and kind help with the statistical analysis methodology adopted in this paper. RSG sincerely thanks the Indian Association for the Cultivation of Science (IACS), Kolkata, for financial and institutional support during the course of this research. The work of TK was supported by Indian Institute of Technology Kanpur.

\bibliography{references}
\end{document}